\documentclass[aps,prb,twocolumn,superscriptaddress,showpacs,amsmath,amssymb,longbibliography]{revtex4-1}
\usepackage[english]{babel}
\usepackage{amsmath}
\usepackage{bm}
\usepackage{graphicx,bbm} 
\usepackage{times}
\usepackage{epsfig} 
\usepackage[colorlinks,linkcolor=blue,citecolor=blue,urlcolor=blue]{hyperref}
\usepackage{color}
\usepackage{soul}
\definecolor{nred} {RGB}{224,0,0}
\definecolor{nblue} {RGB}{28,130,185}
\definecolor{dgreen} {RGB}{78,138,21}
\definecolor{norange}{RGB}{230,120,20}

\begin{document} 
%\title{Delocalization of a particle in a random potential coupled to itinerant bosonic degrees of freedom }
\title{%Dynamics of a particle in a random potential coupled to various bosonic baths  \\
Dynamics of the one--dimensional Anderson insulator coupled to various bosonic baths \\}
%\title{ }
%\title{ }
%\title{}

\author{Janez Bon\v ca}
\affiliation{Faculty of Mathematics and Physics, University of
Ljubljana, SI-1000 Ljubljana, Slovenia}
\affiliation{Jo\v zef Stefan Institute, SI-1000 Ljubljana, Slovenia}
\author{Stuart A. Trugman}
\affiliation{Center for Integrated Nanotechnologies, Los Alamos National Laboratory, Los Alamos, NM}
\author{Marcin Mierzejewski}
\affiliation{Department of Theoretical Physics, Faculty of Fundamental Problems of Technology, Wroc\l aw University of Science and Technology, 50-370 Wroc\l aw, Poland}
%\date
\begin{abstract} 
We study a particle which propagates  in  a one dimensional  strong  random potential  and is coupled to  a bosonic bath. We independently test various  properties of bosons  (hopping term, hard--core effects and generic boson-boson interaction) and show that  bosonic  itineracy is the essential ingredient  governing  the dynamics  of the particle.  Coupling of the particle to itinerant  phonons  or  hard core bosons alike leads to delocalization of the particle  by virtue of a subdiffusive (or diffusive) spread from the initially localized state.  Delocalization remains in effect even when the boson frequency and the bandwidth of itinerant  bosons  remain an order of magnitude smaller than the magnitude of the random potential. When the particle is coupled to localized bosons, its spread remains logarithmic or even sub-logarithmic. The latter result together with the survival probability shows that the particle remains
localized despite being coupled to bosons.\end{abstract}
\pacs{71.23.-k,71.27.+a, 71.30.+h, 71.10.Fd}
\maketitle
%  71.27.+a Strongly correlated electron systems; heavy fermions
%  71.30.+h	Metal-insulator transitions and other electronic transitions
%  71.10.Fd	Lattice fermion models (Hubbard model, etc.)
% 71.23.An Theories and models; localized states
%71.23.-k	Electronic structure of disordered solids
%----------------------------------------------------------------------------------------

\section{Introduction} 
The interplay  between disorder and many-body interactions is a long--standing  problem which is important for the presence of the Anderson localization (AL) \cite{anderson58} in realistic materials. While the problem was recognized many years ago,\cite{ramakrishan1985,fleishman80} recently there has been a significant progress in understanding the physics  of the many-body localization (MBL)  which extends the concept  of AL by  accounting for interactions between the localized particles.\cite{basko06,oganesyan07} The presence of MBL in strongly disordered chains of spinless fermions (or equivalent models) has  consistently been confirmed by various theoretical investigations 
\cite{ZZZ5_1,ZZZ5_2,ZZZ5_3,ZZZ5_4,ZZZ5_7,ZZZ5_8,ZZZ5_9,ZZZ5_10,ZZZ6_1,torres15,torres16,ZZZ5_5,ZZZ5_6,gopal2,Hauschild_2016,herbrych13,imbrie16,steinigeweg15}  and a few experimental 
studies.\cite{kondov15,schreiber15,JJJ2,bordia16,bordia2017_1,smith2016} The many-body interaction is responsible for  several distinctive features of the MBL systems, in particular for the unusually slow  
dynamics.\cite{znidaric08,bardarson12,kjall14,serbyn15,luitz16,ZZZ4_1,ZZZ4_2,ZZZ2_2,agarwal15,gopal15,znidaric16,lastsub,our2016,kozarzewski16,lev14,lev15,barisic16,bonca2017,lastsub,zakrzewski}

The particle localization is not immune against arbitrary many-body interaction and mechanisms which are known to destroy the Anderson insulator may destroy the MBL as well. In particular, the Anderson insulator may be destroyed by the electron--phonon interaction via the so called phonon-assisted hopping.\cite{MOTT1968, emin1975} However,  the insulating state may still survive 
 in the low-temperature regime, as recently suggested in Refs.  \onlinecite{sante2017,yao2017}.
 The phonon-assisted hopping  has been intensively studied and is mostly understood for regular noninteracting bosons.\cite{emin1975} However, already the case of strictly dispersionless  phonons may pose problems especially in one dimensional  (1D) systems.\cite{emin1975}  The role of other bosonic excitations (e.g. magnons) or the boson-boson interaction
remains unexplored. In particular, it is an open problem whether coupling between charge carriers and magnetic excitations \cite{gopal1,peter, su21,su22,su23,bonca2017,mondaini15} may play the same role as the  electron-phonon coupling.  The essential difference between both types of bosons is that
 the energy density  of the magnetic excitations is bounded from above, whereas phonons can in principle absorb arbitrary energy.  
  
  %Level statistics obtained for strongly disordered 1D Hubbard model with spin-1/2 fermions suggests that MBL may arise also in this model for sufficiently strong charge disorder. On the one hand, other results [] suggest absence of full MBL in the Hubbard model in that spin excitations remain delocalized. On the other hand, such situation may not be generic for arbitrary doping since, e.g., for vanishing concentration of electrons both spin and charge degrees of freedom should be localized.  Moreover, a single hole in a disordered t-J model seems also to be delocalized as long as it couples to itinerant magnetic excitations. 

	Here, we study  a single particle in a disordered chain which is coupled to bosons. We aim to establish which properties of the bosonic system are essential  for preserving/destroying the localized state. In particular, we study systems with regular bosons (e.g. phonons) and hard-core (HC) bosons, whereby the latter case should simulate  spin excitations. We compare  results for itinerant and localized/dispersionless  bosons as well as interacting and noninteracting bosons.  We find that itineracy is essential for localization. We show that for sufficiently strong disorder the particle is localized despite coupling to localized hard-core bosons. However, even very small bosonic dispersion destroys localization and leads to a subdiffusive hole propagation, which may eventually turn into the diffusive transport  at extremely long time--scale. In the system of itinerant noninteracting bosons  the particle and energy transport is ballistic. In order to eliminate artifacts originating from this peculiarity of the bosonic subsystem  we consider also a generic case  with boson--boson interaction  when the energy transport within the bosonic subsystem is diffusive. It turns out that the latter interaction  hardly influences propagation of the coupled particle.   Finally,  the transport in the strongly disordered Holstein model with dispersionless regular bosons is shown to be indeed singular since the particle spreads out logarithmically  or sub--logarithmically in time.

\section{Model and method} 

We investigate the Aanderson localization in the one-dimensional model with a single electron  in a random potential  $\epsilon_j \in [-W,W]$ coupled to bosonic degrees of freedom
\begin{eqnarray}
H &=& -t_{0}\sum_{j} \left[c_{j}^{\dagger}  c_{j+1} + \mbox{h.c.}  + \epsilon_j n_j\right] \nonumber\\
&-&g \sum_{j} n_j (b_{j}^\dagger + b_{j}) +  \omega \sum_{j} b_{j}^\dagger  b_{j} \nonumber \\
&+& t_\mathrm{b} \sum_j \left [ b_j^\dagger b_{j+1} + \mbox{h.c.} \right ] \nonumber \\
&+& V_1\sum_j m_jm_{j+1} + V_2\sum_j m_jm_{j+2},
\label{ham_parts}\\
\end{eqnarray}
where $n_j = c_j^\dagger c_j$ represents the electron number operator,    $b_j$ represents either phonon or HC boson, and $m_j=b_j^\dagger b_j$ is the boson number operator. The strength of electron-boson interaction is given by $g$, $\omega$ is the bosonic frequency.  Dispersion  of  otherwise localized bosonic degrees of freedom is introduced via the overlap integral  $t_\mathrm{b}$, while $V_1$ and $V_2$ represent nearest and next nearest  neighbor bosonic interaction strengths.  
We separately consider standard bosons and the HC bosons.  The former case is  relevant for systems where the quantum particle ($c_i$)
is coupled to optical phonons ($b_i$) with frequency $\omega$. Then,  $[b_i,b^{\dagger}_j ]=\delta_{ij}$ and, in principle, the density of bosonic excitations
may be arbitrarily large.  Choosing $V_1=V_2=0$ one obtains the standard Holstein model.
The results for HC bosons simulate  coupling to spin fluctuations. In this case, the energy spectrum is bounded from above since
there is at most one HC boson per site,  $b^{\dagger}_i b^{\dagger}_i =0$. This restriction shows up  in specific commutation relations $[b_i,b^{\dagger}_j ]=\delta_{ij}(1-2 b^{\dagger}_i b_i)$ for the latter operators. 
We perform calculations for one-dimensional chains of various length sizes  with open boundary conditions. 
%%M% We have limited our calculations to odd chains. 
We perform time-evolution using Lanczos based technique and use the limited functional Hilbert space  (LFHS) first developed in Ref.~\onlinecite{bonca1999}. In the Appendix~\ref{LFS} we give a brief overview of the method. Such approach has successfully been applied  to studies on the real--time dynamics of $t$--$J$ and Holstein models.\cite{JJJ5,JJJ6,JJJ7,JJJ8,JJJ9,bonca2017,vidmar_11,kogoj16,kogoj_prl}
This method enabled calculations on larger chains with open boundary conditions where  the maximal distance between the electron and boson excitation is given by   $N_h$.  
When the numerical calculations are carried out for systems of size $L$, the finite--size analysis usually consists in fitting the results by a function which is  linear in $1/L$. In the present  approach, we find the best fits which are linear in $1/N_h$ and then we take the limit   $N_h \rightarrow \infty$ for the fitting function.  

\section{Numerical results}
We start the time evolution from  a random configuration of bosonic degrees of freedom and well defined original position of the  coupled particle.
%In addition we compute static expectation values of various physical quantities for eigenstates in the middle of the energy band using ARPAC Lanczos techniques. 
We typically take $1400$ realizations of the disorder. In the case of HC boson (HCB) such  choice of the initial state  represents propagation at infinite temperature. This is not the case for the Holstein model due to unlimited number of phonon degrees of freedom. In the later case the temperature
of bosonic subsystem is quite elevated but still finite. In the Appendix \ref{apc} we discuss how results depend on the initial state of the bosonic bath. We measure time in units of $[\hbar/t_0]$, in addition for simplicity we set in all cases $t_0=\omega=g=1$.

In order to investigate the dynamics of the charge carrier we calculate the particle  density 
\begin{equation}
\rho_j=\langle \psi(t) |n_{j}  |\psi(t) \rangle_\mathrm{ave},
\label{rho}
\end{equation}
where the index "$\mathrm{ave}$" signifies that expectation values  have  been averaged over different random realisations of ${\epsilon_j}$. 
Since  the density is normalized, $\sum_j \rho_j=1$,  we also    define  the mean square deviation of the hole distribution \cite{robin2017}

\begin{equation}
%\sigma^2(t)=\sum_i i^2 \rho_i(t) -\left[\sum_i i \rho_i(t) \right]^2.
\sigma^2=\sum_j j^2 \rho_j -\left[\sum_j j \rho_j \right]^2.
\label{vari}
\end{equation}

\begin{figure}[!htb]
\includegraphics[width=0.9\columnwidth]{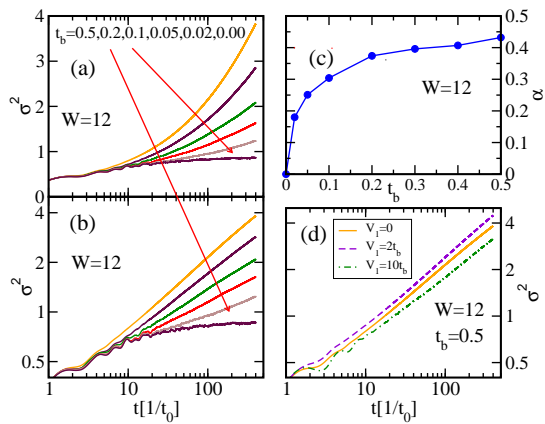}
\caption{ The mean square deviation $\sigma^2(t)$  for different values of $t_\mathrm{b}$ for the case when the particle is coupled to HCB's using a) semi--log plot and b) log--log plot; c) fitting exponents $\alpha$ vs. $t_\mathrm{b}$ extracted   from $\sigma^2(t)=A t^\alpha$.  Fitting was performed in the long-time-limit; d)  log--log plot of $\sigma^2(t)$ of the HCB model   using different values of nearest  and next nearest interaction $V_1$ and $V_2$, respectively.  In all cases we have used $\omega = g = 1$, $W=12$,  and $N_h=20$.  }
\label{fig1}
\end{figure}

We start by presenting results for the HCB  model. In   Fig.~\ref{fig1} we present the time evolution of $\sigma^2$ at   large disorder $W=12$. In the case of localized  HCB's, {\it i.e.} when $t_\mathrm{b}=0$, $\sigma^2(t)$ approaches a constant, indicating particle localization. In contrast, even a small value of dispersion $t_\mathrm{b} > 0$ already leads to a power law behavior, {\it i.e.}  $\sigma^2\propto t^\alpha$,   clearly demonstrated as a straight line on the log--log plot, see Fig.~\ref{fig1}(b).  It is also instructive to note that the  time scale when the power law sets in is roughly given by $1/t_\mathrm{b}$ as most clearly observed as a deviation from the straight line in Fig.~\ref{fig1}(b). In Fig.~\ref{fig1}(c)  we display extracted exponents $\alpha(t_\mathrm{b})$. \ They appear to be non--universal and characteristic of a subdiffusive spread of the initially localized particle. Moreover,  in the whole range of $t_\mathrm{b}$ their values remain $\alpha(t_\mathrm{b})<0.5$, which is far below $\alpha=1$ that is  distinctive  for the diffusive spread. Moreover, from our analysis we may extrapolate that $\alpha(t_\mathrm{b}>0)>0$ suggesting  that the particle remains localized only in the   dispersionless  limit when HCB's are strictly localized, {\it i.e.} at $t_\mathrm{b}=0$.  This is perhaps expected from the point of view of variable range hopping theory \cite{mott89}
and Fermi golden rule, which assumes that the bosons created in the inelastic
hopping process spread out to infinity, hence the probability for the reabsorbtion by the electron drops to zero.
Nonzero $t_b$ lets bosons spread out. In contrast,   at  zero $t_b$ they remain in the vicinity of the particle, consequently  an emitted boson can be reabsorbed to reverse the hopping process.

So far we have shown that already  a small amount of dispersion among HCB's leads to a delocalization of a particle in a one-dimensional random potential. This holds true even  when the magnitude of the random potential $W$ by far exceeds the boson frequency $\omega$ and the  bandwidth $\Gamma=4t_\mathrm{b}$. Next we investigate the influence of interactions between HCB's. In Fig.~\ref{fig1}(d) we present results for a fixed value of disorder, at finite value of $t_\mathrm{b}$ but different choices of nearest  and next--nearest interactions, $V_1$ and  $V_2$, respectively. We further fix the value of $V_2=V_1/2$. The reason for the choice of a finite value of $V_2$ is that at $V_2=0$ the system of interacting HCB's with zero coupling to the particle is exactly solvable and shows ballistic energy transport.
One expects that the electron-phonon coupling alone is sufficient to restore the normal diffusive transport in the bosonic subsystem, even for $V_1=V_2=0$.  
It is clearly the case for non-zero concentration of particles. However, this mechanism may not be efficient for the present case of  a single particle which couples to much larger bosonic bath since  
the relevant time--scale for the onset of normal transport may be very long. 
 In comparison to the $V_1=0$ case we observe a slight  increase of $\sigma^2(t)$ at small $V_1=2t_\mathrm{b}$  followed by a decrease with further  increasing of $V_1$ towards $V_1=10 t_\mathrm{b}$. In the latter case we also observe a small decrease of $\alpha$. Interactions  among HCB's have only a small effect on the delocalization processes. The slight  increase in $\sigma^2(t)$ at small values of $V_1$ can be due to lifting  of the degeneracy among many--body HCB states in the presence of  interactions. However, further increasing of $V_1$ may lead to slowing down of the propagation of excitations  in the HCB subspace that seem to be  responsible for the delocalization of the particle.

\begin{figure}[!htb]
\includegraphics[width=0.9\columnwidth]{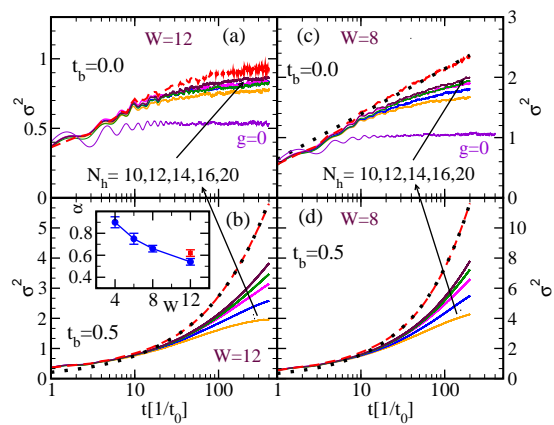}
\caption{  Semi--log plots of $\sigma^2(t)$ of the HCB model for different  sizes of the Hilbert spaces generated by $N_h$. Cases with no dispersion,   $t_\mathrm{b}=0$, are presented in  a) and c) for two distinct values of disorder $W=12$ and 8, respectively. Results for $t_\mathrm{b}=0.5$  are shown as well for two distinct values of $W$ in  b) and d).  Note also substantially different scales used to present results with or without dispersion.    Dashed (red) lines represent results after finite--size scaling analysis. Dotted line in c) represents a fit to the form $\sigma^2(t)=A + B\log(t). $ Dotted lines in b) and d) represent fit to the form $\sigma^2(t)=At^\alpha$. Thin (violet in colour) line in a) and c) represents evolution of $\sigma^2(t)$   for a free particle, {\it i.e.} $g=0$, that is subject to  Anderson's localization. Inset in  b)  displays  exponents $\alpha$ (circles) extracted from  finite--size scaled results at different values of disorder $W$. A singular square represents result for the Holstein model with identical parameters as the HCB one. Other parameters of the model were  $\omega=g=1$ and $V_1¤=V_2=0$.    }
\label{fig2}
\end{figure}

We have tested the validity of our findings with regard to  finite--size effects as well as regarding the effect of limited functional Hilbert spaces used in our calculations. The size of the LFHS exponentially depends on the parameter $N_h$. We refer the reader for a more precise explanation of the meaning of $N_h$ to the Appendix~\ref{LFS} as well as to the original publication in Ref.~\onlinecite{bonca1999}.  Here we only note  that $N_h$ represents the maximal length that the particle travels from its original position, while the maximal number of HCB's is given by $N_h/2$. In Fig.~\ref{fig2} we show results for two different values of disorder,  obtained with different Hilbert spaces.  When HCB's are localized, {\it i.e.} for  
$t_\mathrm{b}=0$,  the particle also remains localized,  see Figs.~\ref{fig2}(a) and (c),   even after the finite--size analysis.  In particular, at $W=8$ we observe a logarithmic  increase of $\sigma^2(t)$, characteristic for MBL systems \cite{our2016,peter2017an}, while at yet stronger disorder, $W=12$, we observe a tendency towards the saturation similar to the case of noninteracting particle (see curve for $g=0$).   However,  in contrast to the noninteracting system,  strict saturation does not arise within the accessible time--window and the extremely slow dynamics resembles the MBL systems, rather than noninteracting AL.

In contrast, in the case of itinerant HCB's, that is at finite dispersion  $t_\mathrm{b}=0.5$,   we observe subdiffusion, see Figs.~\ref{fig2}(b) and (d). Dashed lines in all cases represent results obtained using finite--size scaling analysis.  For finite dispersion, we have obtained nearly perfect fits, presented with dotted lines,  to the analytical form $\sigma^2(t)=A t^\alpha$. We have performed a similar analysis as well for smaller values of $W=6$ and 4, not shown. In the inset of Fig.~\ref{fig2}(b) we show extracted $\alpha$'s that are  increasing towards $\alpha=1$ as the disorder decreases. Due to increasing finite--size effects we were unable to reliably investigate systems with $W<4$.

\begin{figure}[!htb]
\includegraphics[width=0.9\columnwidth]{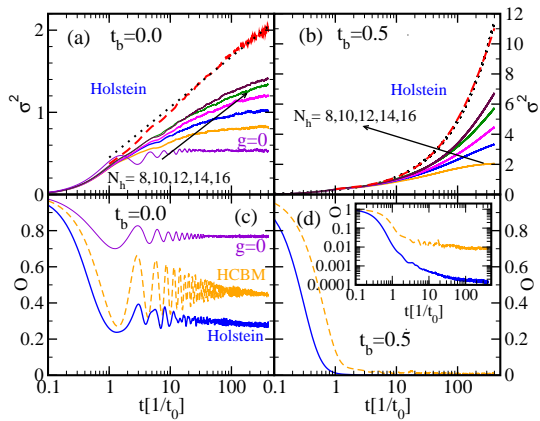}
\caption{  a) and b)  $\sigma^2(t)$ of the Holstein  model for different  sizes of the Hilbert spaces generated by $N_h$  and $W=12$. The cases with no dispersion,   $t_\mathrm{b}=0$, are presented in  a) while results for $t_\mathrm{b}=0.5$  are shown b).  Dashed (red) lines represent results after finite--size scaling analysis.   Dotted line in a)   represents a fit to the form $\sigma^2(t)=A + B\log(t). $ Dotted line in b)  represent fit to the form $\sigma^2(t)=At^\alpha$. Overlaps  $\cal O$  are shown in c) ($t_b=0$) and d) ($t_b=0.5$) for the Holstein model with  full lines and the HCBM with dashed lines. 
Thin   lines (violet in colour) in a) and c) represent evolution of $\sigma^2(t)$  and $\cal O$, respectively,  for a free particle, {\it i.e.} at $g=0$, that undergoes  Anderson's localization.
The inset in d) shows the same data as in d) but using log-log scale.  Other parameters of the model were in all cases  $\omega=g=1$,  $W=12$, and $V_1=V_2=0$.   }
\label{fig3}
\end{figure}

We next present results for the Holstein model. Due to unlimited phonon degrees of freedom we had to limit our calculations to a  maximal number of phonons, given by $N_h$.  Similarly to the HCB model case, $N_h$  represents also the maximal distance that the particle travels from the origin,  while $N_h-1$ is the maximal distance between the particle and a single  phonon excitation.  We start the time evolution from  an initial  random configuration of phonon degrees of freedom and well defined initial  position of the particle. 
We present results in Fig.~\ref{fig3} for a single set of parameters, {\it i.e.} $W=12$ as well as at fixed $g=1$. The discussion of the influence of increasing coupling constant from weak towards strong coupling limit for the Holstein model   is presented in Appendix~\ref{scl}. In the case of localized phonons, {\it i.e.} $t_\mathrm{b}=0$, we observe slow, logarithmic increase of $\sigma^2(t)$, see Fig.~\ref{fig3}(a). 
Since  we have used identical parameters as in the HCB model, Figs.~\ref{fig2}(a) and ~\ref{fig3}(a) provide direct comparison between the models. While in the case of the HCB model  $\sigma^2(t)$ shows signs of saturation or at most sub--logarithmic growth, 
we observe a clear logarithmic growth   when the particle is coupled to regular bosons (phonons). Moreover,  its spread is enhanced in comparison to the HCB case and displays quantitatively distinct behavior from the noninterracting case at $g=0$. A similar comparison is found as well in the case of finite dispersion. Coupling to itinerant phonons again  leads to a subdiffusive growth of $\sigma^2(t)$, see Fig.~\ref{fig3}(b) with an exponent $\alpha=0.62$ that is about 10\% larger than in the case of the HCB model, for visual comparison see also the inset of Fig.~\ref{fig2}(b).  However, one cannot exclude that in this model there exists 
small, albeit nonzero diffusion constant.  Then, the subdiffusion would be a transient effect since the spread due to normal diffusion will dominate at sufficiently
long time. Suppression of transport in the case when particle is coupled to marginally localized phonons has recently  been demonstrated in Ref. \onlinecite{altman2016} for the
low--temperature regime.

Finally, we investigate  the survival probability defined as  the overlap of the many--body wave function $\vert \psi(t)\rangle$ with the initial one  $\vert \psi(0)\rangle$.\cite{heyl2014,torres15,peter2017} Namely, we compute 
\begin{equation}
{\cal O}(t) = \vert\langle \psi(0)\vert \exp(-iHt) \vert\psi(0)\rangle\vert^2_{\mathrm{ave}}.
\end{equation}
It measures the probability for finding the system still in the initial state $\vert \psi(0)\rangle$ at time $t$.\cite{torres15} In the case of localized bosons, $t_\mathrm{b}=0$ presented in Fig.~\ref{fig3}(c), ${\cal O}(t)$ approaches a constant in the long--time limit. In addition,  well defined  oscillations with a frequency $\omega \sim 2$ are observed at moderate times that are more pronounced in the HCBM case. They signal transitions among only a few states. 
The specific value of the frequency originates from the disorder--averaging and indicates that the charge dynamics is well restricted to the neighboring sites.\cite{kozarzewski16}  In contrast, in the case of itinerant bosons, Fig.~\ref{fig3}(d), 
${\cal O}(t\to \infty )\to 0$ while oscillations are strongly overdamped.   The survival probability turned out to be very useful in the studies concerning the many--body localization,\cite{torres15,peter2017} where ${\cal O}(t)$ decays exponentially with the system size  $L$,\cite{peter2017}     ${\cal O}(t \gg 1) \sim \exp(- a L)$.   The latter holds true in the MBL as well as in the ergodic regimes, whereby the parameter $a$ in the localized system is much smaller than in the ergodic case. In the present studies, we have found a clear exponential decay with $N_h$ (quantity equivalent to $L$) only for systems with itinerant bosons, see the discussion in the Appendix~\ref{FSS}. In contrast,   in systems with localized bosons the dependence of ${\cal O}(t \gg 1)$ on the system size is rather small.  
The survival probability  is constructed in terms of the many-body wave function of the total system.
Then, the finite value of ${\cal O}(t)$ in the long--time limit and weak $N_h$--dependence do not only indicate localization of the particle but also freezing of the initial distribution of bosons with $t_b=0$.

\section{Conclusions} 
We have studied the time evolution of a particle in a strong random potential coupled to localized or itinerant bosonic degrees of freedom. The study was based on a Holstein--like model in one  dimension. Two types of bosons, {\it i.e.} hard core bosons and phonons were used in our  study.  The main motivation to study  hard core bosons was on the one hand their similarity to spin degrees o freedom and on the other their limited degrees of freedom that allowed  studying larger system sizes.  The coupling of the particle to itinerant bosons, HCB's and phonons alike,  leads to delocalization  by virtue of a subdiffusive spread from the initially localized state.  Even more surprisingly,  delocalization remains in effect even when the boson frequency and the bandwidth of itinerant  bosons  remain an order of magnitude smaller than the magnitude of the random potential. From among all the discussed properties of the bosonic bath, the itineracy of bosons plays the crucial role for the dynamics of the interacting particle. 

We expect  for dispersive standard bosons that the subdiffusive transport may be a long--lasting but still  a transient phenomena. On a very long time--scale, the particle dynamics should be similar
to that discussed in Ref. \onlinecite{gopal2}  where, instead of phonons, the quantum particle is coupled to a classical noise. However for strongly disordered systems, the time--scale corresponding to the onset of standard diffusion is beyond the reach of direct numerical calculations for many--body quantum systems. Even more challenging question concerns the asymptotic dynamics of particle
coupled to the hard--core bosons.  
% Adding comment from stuart %%%%%%

{\it Acknowledgments.}
J.B. acknowledges the financial support from the Slovenian Research Agency  (research core funding No. P1-0044) and M.M. acknowledges support by the project 2016/23/B/ST3/00647 of the National Science Centre, Poland. S.A.T. acknowledges support from CINT. This work was performed, in part, at the Center for Integrated Nanotechnologies, a U.S. Department of Energy, Office of Basic Energy Sciences user facility.

\appendix

\section{Generator of Limited Functional Hilbert Space }  
\label{LFS}
We only give a short description of the main parts of the method. More details can be found in the original work, Ref.~\onlinecite{bonca1999}. We  choose the   generator of the Limited Functional Hilbert Space (LFHS) that consists of two off-diagonal parts of the Hamiltonian  in Eq.~(1) of the main text,
\begin{eqnarray}
{\cal O}_1&=& \sum_{j} n_j (b_{j}^\dagger + b_{j}) \\
{\cal O}_2&=& \sum_{j} c_{j}^{\dagger}  c_{j+1} + \mbox{H.c.}  \quad . 
\end{eqnarray}
The generating algorithm starts from a particle at a given position, {\it e.g.} $j=0$, in a vacuum state of boson excitations, 
$\vert \psi^{(0)}\rangle = c^\dagger_{0\sigma} \vert  0\rangle$ where $\vert 0\rangle$ represents vacuum for the particle as well as boson excitations.   We then apply the generator of basis states  $N_h$-times to generate the LFHS:
\begin{equation} \label{lfs}
\left\{|\psi^{(l)} \rangle \right\}= \left( {\cal O}_1+{\cal O}_2\right)^{l}|\psi_{(0)} \rangle,
\end{equation}
for $l=0,...,N_h$. We thus generate a Limited Functional Hilbert Space spanned by states of the following form
\begin{equation}
\vert\psi\rangle = \vert j;\dots,n_{j-1},n_j,n_{j+1},\dots\rangle
\end{equation}
where $j$ represents the particle coordinate, while there are $n_m$ bosons on site $m$. In the HCB case,  $n_j\in \{ 0,1 \}$ while  for phonons $n_j\in \{ 0,\dots,N_h \}$. The 
limited  functional  Hilbert space that we construct is not a standard one where bosonic degrees of freedom would be distributed uniformly on the lattice irrespective  to the particle position. Our approach  adds basis states more efficiently than
some other methods.  In the case of generating phonon degrees of freedom  a basis state is included if it can be
reached using $N_b$ phonon creation operators and $N_t$ particle
hops in any order with $N_b + N_t\leq N_h$ . For a given $N_h$ , there
is a basis state with $N_h$ phonon quanta on the same site as the particle and no phonon excitations elsewhere. The particle can hop maximally $N_h$ sites away from its original position, but then there is no boson nor phonon quanta in the system. In the HCB case the maximal number of boson quanta is $N_h/2$.  It is achieved by successive process where a HCB is created on site-$j$ followed by a jump to site $j+1$. In the case of LFHS we impose open boundary conditions. After completing generation of LFHS we time evolve the wave function using the Hamiltonian in Eq.~(1) of the main text  while taking  advantage of the  standard Lanczos-based diagonalization technique. Sizes of LFHS for the HCB model span from $N_\mathrm{st}\sim 10^3$ for $N_h=10$ up to  $2 \times 10^5 $ for the largest $N_h=20$ used in our calculations. 
Sizes of LFHS for the Holstein  model span from $N_\mathrm{st}\sim 10^3$ for $N_h=8$ up to  $5 \times 10^5 $ for the largest $N_h=16$.  To achieve sufficient  accuracy of time propagation, we have used time-step-size $\Delta t=0.02$ and  performed up to $2\times 10^4 $ time steps. In addition we have sampled over $10^3$ different realizations of disorder  $\epsilon_i$. 

The main advantage of LFHS over the exact diagonalization approach  is to significantly reduce the dimension of the Hilbert space.
The method has been successful in computing properties of the driven Holstein polaron,\cite{vidmar_11} dissociation of a driven bipolaron,\cite{golez_12} relaxation dynamics and thermalization properties of a highly excited polaron,\cite{kogoj16,kogoj_prl,JJJ9} as well as static and dynamic properties  \cite{JJJ8} and non-equilibrium dynamics \cite{JJJ7,JJJ6,JJJ5} of correlated electron systems.

\section{Initial state of the bosonic bath}  
\label{apc}

\begin{figure}[!htb]
\includegraphics[width=0.9\columnwidth]{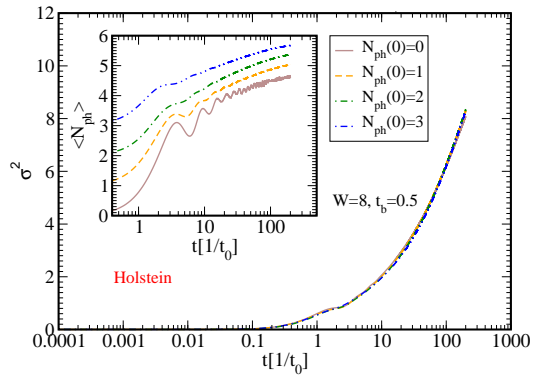}
\caption{ $\sigma^2$ vs. $t$ for the Holstein model. $N_{\mathrm{ph}}(0)$ the  number of phonon excitations at $t=0$. Inset represents $\langle N_{\mathrm{ph}} \rangle (t)$. In both  cases we have used $\omega = g = 1$, $t_b=0.5,$ and $W=8$.  }
\label{figS3}
\end{figure}
 We test the dependence of $\sigma^2$ on the initial  state of the bosonic bath in the case of the Holstein model. In Fig.~\ref{figS3} we present $\sigma^2(t)$ obtained by starting the time propagation from  random  initial  states characterized by different  total number of bosonic excitations $\langle N_{\mathrm{ph}} \rangle  \in \{0,1,2,3\}$, where $N_{\mathrm{ph}}  = \sum_j b_j^\dagger b_j$. In the inset of Fig.~\ref{figS3} we also follow the time evolution of $\langle N_{\mathrm{ph}} \rangle (t)$. Different initial states  in the long-time evolve towards  distinct bosonic states, nevertheless, the spread of the initially localized particle   $\sigma^2(t)$ remains nearly independent on the state of the bosonic subspace.
 
In the case of thermal equilibrium, 
different values of $\langle N_{\mathrm{ph}} \rangle$ correspond to different temperatures. It should be noted, however, that the system under consideration is initially not in the thermal state.
Fig.\ref{fig1}d in the main text  shows that the spread of the quantum particle in HCBM is weakly modified by the boson--boson interaction even for very strong potentials $V_1$ and $V_2$. Since  the latter interaction  should lead to a rather fast thermalization of the bosonic bath we come to conclusion that non--thermal initial state of the bosonic bath does not influence the spread of particle, at least not on a qualitative level.

\section{Strong coupling limit}  
\label{scl}
\begin{figure}[!htb]
\includegraphics[width=0.9\columnwidth]{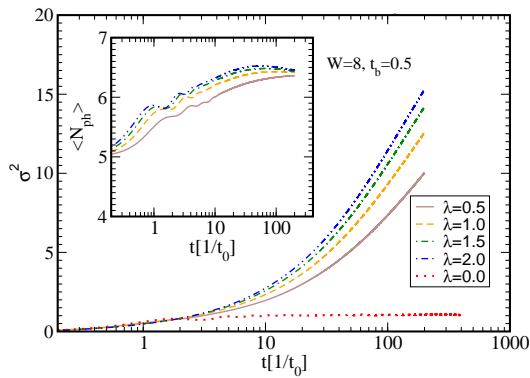}
\caption{ $\sigma^2$ vs. $t$ for the Holstein  model for different coupling strengths $\lambda$.  Inset represents $\langle N_{\mathrm{ph}} \rangle (t)$. In both  cases we have used $\omega = 1$, $t_b=0.5$, and $W=8$. }
\label{figS4}
\end{figure}

 Here we explore  the influence of the coupling constant $g$ on the dynamics of the particle. We first introduce the dimensioneless coupling constant $\lambda = g^2/2\omega t_0$. It is well known that $\lambda\sim 1$ represents the transition point between the weak-coupling regime for $\lambda \lesssim 1$ and the  strong coupling one $\lambda \gtrsim 1$. In  the latter the polaron effective mass scales approximately as $m^*\propto \exp(g^2)$. The naive expectation is then that by increasing $\lambda$ the particle would become nearly localised  due to the exponentially increased $m^*$. In contrast, as shown in Fig.~\ref{figS4}, the increase of $\lambda$ leads to a monotonous increase of $\sigma^2$. It should be noted that during the time evolution  the system  evolves through highly excited states while the concept of a polaron with a large effective mass is a ground state phenomena. Emission and subsequent reabsorption of phonons represents the main mechanism for delocalisation of the particle in a random potential.  For comparison we also include result for $\lambda=0$ that shows Anderson's localisation.

\begin{figure}[!htb]
\includegraphics[width=0.9\columnwidth]{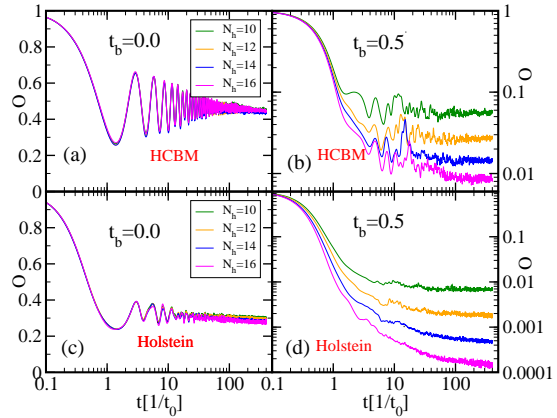}
\caption{ ${\cal O}(t)$ as given in Eq.~(5)  of the main text using different system sizes as given by $N_h$,  for the HCBM in (a) and (b) and the Holstein model in (b) and (d). In all cases we have used $\omega = g = 1$, $W=12$.  }
\label{figS1}
\end{figure}

\section{Finite-size scaling of the survival probability  }  
\label{FSS}
In Fig.~\ref{figS1} we present finite-size scaling of the survival probability ${\cal O}(t)$, as defined in Eq.~(5) of the main text,  in the limit of large disorder, $W=12$. In the case of localized bosons, {\it i.e.} at $t_\mathrm{b}=0$, see Figs.~\ref{figS1}(a) and (c),  we observe near complete overlap of results obtained using system sizes ranging from  $10^3$ states in cases of $N_h=10$ through $10^6$ in the case of  $N_h=16$. In contrast, in the case of itinerant bosons, for $t_\mathrm{b}=0.5$, we observe a substantial  $N_h$-dependence of  ${\cal O}(t>>1)$ in both models, see Figs.~\ref{figS1}(b) and (d). Note also that in contrast to the previous case, latter results are presented on a log-log scale. Then, almost equally spaced flat sections of  ${\cal O}(t>>1)$ obtained for $N_h=10,12,14,...$ indicate that the survival probability scales exponentially with the system size $L$  that is in our method given by $L\sim N_h$, {\it i.e.} ${\cal O}(t>>1)\propto \exp(-aN_h)$. 

\begin{figure}[!htb]
\includegraphics[width=0.9\columnwidth]{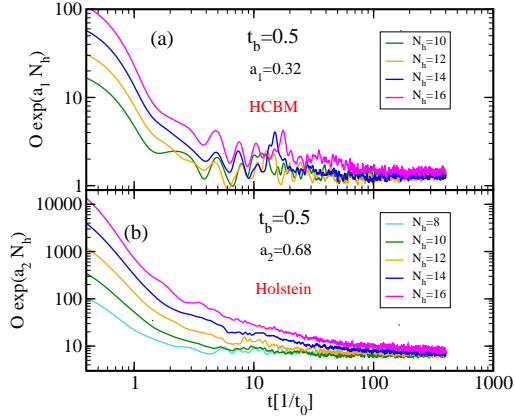}
\caption{ Scaled survival probability ${\cal O}(t) \exp(a_iN_h)$ for the case when $t_\mathrm{b}=0.5$ for the two models as indicated in the figure.  In both  cases we have used $\omega = g = 1$, $W=12$.  }
\label{figS2}
\end{figure}

In order to obtain a more quantitative picture of the above mentioned exponential scaling  we present in Fig.~\ref{figS2} ${\cal O}(t)\exp(a_iN_h)$ where $i=1,2$ for the two models under consideration. A nearly perfect scaling is observed for $t>>1$. 

It is beneficial  to stress two important properties of the localized state in systems with  $t_b=0$.  On the one hand, there is an extremely long-time scale 
which governs the particle dynamics for $t \gtrsim 10^2$ as it is clearly visible in Figures  2(a) and 3(a) in the main text. 
Such slow dynamics is  characteristic for MBL systems,\cite{our2016,peter2017an} whereas  it does not arise in the Anderson insulators ($g=0$) where the spreading of particle saturates already at
$t \sim 10$.  On the other hand, the  survival probability does not show any clear exponential decay with the system size, as it is the case in the  MBL.\cite{peter2017}  The survival probability in the studied electron-phonon system with $t_b=0$ resembles rather the projection of single--particle wave functions $\langle \psi_{sp}(0)  |\psi_{sp}(t) \rangle$  in the Anderson insulators, which  is for the particle under consideration at zero electron-phonon coupling presented in Fig.~3(c) of the main text. 

%----------------------------------------------------------------------------------------
%\bibliographystyle{apsrev4-1}
\bibliography{references}
\end{document}